\documentclass[final,3p,twocolumn,times]{elsarticle}

\usepackage{graphicx}
\usepackage{amsmath}
\usepackage{amssymb}
\usepackage{amsthm}
\usepackage{bm}
\usepackage{color}
\usepackage{stmaryrd}
\usepackage{mathrsfs}
\usepackage{array}
\usepackage{algorithm}
\usepackage{algpseudocode,algpseudocode}
\usepackage{multirow}
\usepackage{url}
\usepackage{lineno}
\usepackage{mdframed}

\SetSymbolFont{stmry}{bold}{U}{stmry}{m}{n}

\numberwithin{equation}{section}
\newcolumntype{P}[1]{>{\centering\arraybackslash}p{#1}}

\newtheorem{assumption}{Assumption}
\AfterEndEnvironment{assumption}{\noindent\ignorespaces}

\journal{Forces in Mechanics}

\begin{document}

\title{Numerical investigation of abdominal aortic aneurysm hemodynamics using the reduced unified continuum formulation for vascular fluid-structure interaction}

\author[1]{Ingrid S. Lan}
\ead{ingridl@stanford.edu}

\author[2,3]{Ju Liu\corref{cor1}}
\ead{liuj36@sustech.edu.cn}

\author[4]{Weiguang Yang}
\ead{wgyang@stanford.edu}

\author[1,4,5]{Alison L. Marsden}
\ead{amarsden@stanford.edu}

\cortext[cor1]{Corresponding author}
\address[1]{Department of Bioengineering, Stanford University, Stanford, CA 94305, USA}
\address[2]{Department of Mechanics and Aerospace Engineering, Southern University of Science and Technology, Shenzhen, Guangdong 518055, P.R China}
\address[3]{Guangdong-Hong Kong-Macao Joint Laboratory for Data-Driven Fluid Mechanics and Engineering Applications, Southern University of Science and Technology, Shenzhen, Guangdong 518055, P.R. China}
\address[4]{Department of Pediatrics (Cardiology), Stanford University, Stanford, CA 94305, USA}
\address[5]{Institute for Computational and Mathematical Engineering, Stanford University, Stanford, CA 94305, USA}

\begin{abstract}
We recently demonstrated the reduction of the unified continuum and variational multiscale formulation to a computationally efficient fluid-structure interaction (FSI) formulation via three sound modeling assumptions pertaining to the vascular wall. Similar to the coupled momentum method introduced by Figueroa et al., the resulting semi-discrete formulation yields a monolithically coupled FSI system posed in an Eulerian frame of reference with only a minor modification of the fluid boundary integral. To achieve uniform second-order temporal accuracy and user-controlled high-frequency algorithmic damping, we adopt the generalized-$\alpha$ method for uniform temporal discretization of the entire coupled system. In conjunction with a fully consistent, segregated predictor multi-corrector algorithm preserving the block structure of the incompressible Navier-Stokes equations in the implicit solver's associated linear system, a three-level nested block preconditioner is adopted for improved representation of the Schur complement. In this work, we apply our reduced unified continuum formulation to an appropriately prestressed patient-specific abdominal aortic aneurysm and investigate the effects of varying spatial distributions of wall properties on hemodynamic and vascular wall quantities of interest.
\end{abstract}


\maketitle

\section{Introduction}
\label{sec:introduction}
The unified continuum and variational multiscale (VMS) formulation for fluid-structure interaction (FSI) was recently developed using the Gibbs free energy as the thermodynamic potential \cite{Liu2018}. The resulting pressure primitive variable formulation not only recovers important continuum models including viscous fluids and visco-hyperelatic solids, but is also well-behaved in both compressible and incompressible regimes, in contrast to conventional formulations based on the Helmholtz free energy. Restricting our attention to vascular FSI, we recently reduced this unified continuum formulation in the arbitrary Lagrangian-Eulerian (ALE) description by adopting three modeling assumptions for the vascular wall, namely the infinitesimal strain, thin-walled, and membrane assumptions \cite{Lan2021}. The resulting FSI formulation, referred to as the \textit{reduced unified continuum (RUC) formulation}, achieves convenient monolithic coupling of the fluid and solid sub-problems in an Eulerian frame through a modification of the fluid boundary integral on the vessel wall. Despite its ostensible similarity to the semi-discrete formulation in the coupled momentum method \cite{Figueroa2006}, our underlying derivation relies only on a conforming mesh rather than an assumed fictitious body force in the solid sub-problem to enable the fluid-solid coupling. Furthermore, uniform temporal discretization of the entire FSI system is achieved via the generalized-$\alpha$ scheme, in which velocity and pressure are concurrently collocated at the intermediate time step for second-order temporal accuracy. This is in contrast to the dichotomous approach commonly adopted in the computational fluid dynamics (CFD) and FSI communities, which we recently found to yield only first-order temporal accuracy for pressure \cite{Liu2020a}. Without loss of consistency, a segregated predictor multi-corrector algorithm is designed to preserve the same block structure as for the incompressible Navier-Stokes equations in the implicit solver's associated linear system. Exploiting this preserved block structure, we then design a nested block preconditioner to attain improved representation of the Schur complement and thus robustness for cardiovascular simulations involving contributions of widely varying orders of magnitude, including convection, diffusion, vascular wall stiffness, and coupled lumped parameter networks representing upstream and/or downstream circulatory components.

Significant interest in the literature lies in using the time-averaged wall shear stress (TAWSS) and oscillatory shear index (OSI) and combinations thereof, such as thrombus formation potential, as hemodynamic predictors of intraluminal thrombus formation and rupture of abdominal aortic aneurysms (AAAs) \cite{Yazdani2017, Doyle2014, DiAchille2014, Wilson2013, Rissland2009, DiMartino2001}. Nonetheless, few computational FSI studies can be found for patient-specific AAAs, likely a consequence of the paucity of spatially resolved patient-specific material properties, as well as the large discrepancies in material behavior reported by in vivo studies \cite{vantVeer2008, Wilson2003} and in vitro uniaxial \cite{Raghavan2000, DiMartino2006, Xiong2008, Reeps2013} and biaxial \cite{VandeGeest2006} tensile testing. In this work, we adopt the RUC formulation for a comparison of AAA hemodynamics using the maximum tangential modulus reported by Vorp et al. from either uniaxial \cite{DiMartino2006} or biaxial \cite{VandeGeest2006, Ferruzzi2011} tensile testing. Importantly, practical cardiovascular simulations for clinical applications require additional modeling techniques to address the nonzero stress state of the vascular wall at imaging and the assignment of variable wall properties. We therefore use a fixed-point iteration algorithm to determine the tissue prestress and a centerline-based approach to assign the local wall thickness.

\section{Numerical formulation}
\label{sec:formulation}
In this section, we first present the governing equations for the FSI problem, beginning with the unified formulation and arriving at the reduced semi-discrete formulation via three assumptions for the vascular wall. We subsequently outline novel aspects of our solution strategy, including the temporal discretization, segregated predictor multi-corrector algorithm, and nested block preconditioner. 

\subsection{Strong-form FSI problem}
We consider a domain $\Omega \subset \mathbb R^3$, which admits a non-overlapping subdivision $\overline{\Omega} = \overline{\Omega^f \cup \Omega^s}$, $\emptyset = \Omega^f \cap \Omega^s$, in which $\Omega^f$ and $\Omega^s$ represent the sub-domains occupied by the fluid and solid materials, respectively. The fluid-solid interface is a two-dimensional manifold denoted by $\Gamma_I$ with unit outward normal vectors $\bm n^f$ and $\bm n^s$ relative to $\Omega^f$ and $\Omega^s$, respectively, such that $\bm n^f = - \bm n^s$.

Under the isothermal condition, the governing equations for the solid material are posed in $\Omega^s$ as follows,
\begin{align}
\label{eq:ela_kinematics}
& \bm 0 = \frac{d \bm u^s}{d t} - \bm v^s, \\
\label{eq:ela_mass}
& 0 = \beta^s_{\theta}(p^s) \frac{dp^s}{dt} + \nabla \cdot \bm v^s, \\
\label{eq:ela_mom}
& \bm 0 = \rho^s(p^s) \frac{d \bm v^s}{d t}  - \nabla \cdot \bm \sigma^s_{\mathrm{dev}} + \nabla p^s - \rho^s(p^s) \bm b^s,
\end{align}
where $\bm u^s$, $\bm v^s$, $p^s$, $\rho^s$, and $\bm b^s$ are, respectively, the solid displacement, velocity, pressure, density, and the body force per unit mass; $\beta^s_{\theta}$ is the isothermal compressibility coefficient; and $\bm \sigma^s_{\mathrm{dev}}$ is the deviatoric part of the Cauchy stress.

\begin{assumption}
\label{assumption:small_strain}
The solid deformation is small enough such that the infinitesimal strain theory is valid.
\end{assumption}
Under the above assumption, we do not distinguish between the reference and current frames, allowing the total and partial time derivatives in \eqref{eq:ela_kinematics}-\eqref{eq:ela_mom} to be used interchangeably. Furthermore, $\beta_{\theta}(p^s) = 1/\kappa^s$, where $\kappa^s$ is the solid bulk modulus, and \eqref{eq:ela_mass} can be integrated to yield $p^s = - \kappa^s \nabla \cdot \bm u^s$ \cite{Liu2019}.

While the fluid sub-problem in an ALE formulation is indeed posed on a moving domain that tracks the solid deformation, Assumption \ref{assumption:small_strain} guarantees this geometry adherence and renders mesh motion unnecessary. The resulting FSI system thus has the same order of algorithmic complexity as that of CFD. Having eliminated mesh motion, the governing equations for an incompressible Newtonian fluid can be written in the Eulerian frame $\Omega^f$ as follows,
\begin{align*}
& \bm 0 = \rho^f \frac{\partial \bm v^f}{\partial t}  + \rho^f  \bm v^f \cdot \nabla \bm v^f - \nabla \cdot \bm \sigma^f_{\mathrm{dev}} + \nabla p^f - \rho^f \bm b^f, \\
& 0 = \nabla \cdot \bm v^f,
\end{align*}
wherein $\bm v^f$, $p^f$, $\rho^f$, $\mu^f$, and $\bm b^f$ are, respectively, the fluid velocity, pressure, density, dynamic viscosity, and body force per unit mass. $\bm \sigma^f_{\mathrm{dev}}$ and $\bm \varepsilon_{\mathrm{dev}}$ are the deviatoric parts of the Cauchy stress and rate-of-strain,
\begin{align*}
& \bm \sigma^f_{\mathrm{dev}} := 2\mu^f \bm \varepsilon_{\mathrm{dev}}(\bm v^f), \\
& \bm \varepsilon_{\mathrm{dev}}(\bm v^f) := \frac12 \left( \nabla \bm v^f + \left( \nabla \bm v^{f} \right)^T \right) - \frac13 \nabla \cdot \bm v^f \bm I.
\end{align*}

The strong-form FSI problem can be completed with the kinematic and dynamic coupling conditions enforcing the continuity of velocity and traction on $\Gamma_I$, respectively,
\begin{align*}
\bm v^f = \bm v^s, \qquad \bm \sigma^f \bm n^f = -\bm \sigma^s \bm n^s.
\end{align*}

\subsection{Reduced semi-discrete FSI formulation}
\label{subsec:semi-discrete-fsi}
Let $\mathcal S_{\bm v}^s$ denote the trial solution space for the solid velocity; let $\mathcal S_{\bm u}^s$ and $\mathcal V_{\bm u}^s$ denote the trial solution and test function spaces for the solid displacement; and let $\Gamma_h^s$ denote the Neumann part of the solid boundary with traction $\bm h^s$ prescribed. The semi-discrete solid formulation can be stated as follows. Find $\bm y^s_h(t):= \left\lbrace \bm v^s_h(t), \bm u^s_h(t) \right\rbrace \in \mathcal S^s_{\bm v} \times \mathcal S^s_{\bm u}$ such that $\forall \bm w^s_h \in \mathcal V^s_{\bm u}$,
\begin{align*}
& \mathbf B^{s}_{\mathrm{k}}\Big( \dot{\bm y}^s_h, \bm y^s_h \Big) := \frac{d\bm u^s_h}{dt} - \bm v^s_h = \bm 0, && \displaybreak[2] \\
& \mathbf B^{s}_{\mathrm{m}}\Big( \bm w^s_h ;  \dot{\bm y}^s_h, \bm y^s_h \Big) := \int_{\Omega^s} \bm w_h^s \cdot \rho^s \left( \frac{d \bm v_h^s}{d t} - \bm b^s \right) d\Omega \nonumber \\
& \hspace{6mm} + \int_{\Omega^s} \bm \epsilon(\bm w_h^s) : \bm \sigma^s(\bm u_h^s)  d\Omega - \int_{\Gamma_h^s} \bm w_h^s \cdot \bm h^s d\Gamma = 0.
\end{align*}

\begin{assumption}
\label{assumption:thin-wall}
$\Omega^s$ is thin in one direction and can thus be parameterized by $\Gamma_I$ and a through-thickness coordinate in the unit outward normal direction.
\end{assumption}
Given this thin-walled assumption, we can parameterize $\bm x \in \Omega^s$ with $\chi \in \Gamma_I$ in the following form,
\begin{align*}
\bm x(\bm \xi) = \bm x(\xi,\eta,\zeta) := \bm \chi(\xi,\eta) + \zeta h^s(\xi,\eta) \bm n^f,
\end{align*}
where $\xi$ and $\eta$ are the in-plane parametric coordinates, $h^s$ is the wall thickness, $\zeta \in (0,1)$ is the through-thickness parametric coordinate, and the unit outward normal $\bm n^f$ can be determined through the following relations,
\begin{align*}
& \bm n^f = \frac{\bm e_{\xi} \times \bm e_{\eta}}{\| \bm e_{\xi} \times \bm e_{\eta} \|},\quad \bm e_{\xi} := \frac{\partial \bm \chi}{\partial \xi} / \left \|\frac{\partial \bm \chi}{\partial \xi} \right\|, \quad \bm e_{\eta} := \frac{\partial \bm \chi}{\partial \eta} / \left \|\frac{\partial \bm \chi}{\partial \eta} \right\|.
\end{align*}
For any fixed $\zeta$, the surface defined by this parameterization of $\Omega^s$ is a lamina, and the coordinate transformation from the global coordinates $\bm x$ to the local lamina coordinates $\bm x^l$ is then given by $\bm x^l = \bm Q \bm x$ with the rotation matrix
\begin{gather*}
\bm Q := 
\begin{bmatrix}
\bm e_1^l & \bm e_2^l & \bm e_3^l
\end{bmatrix}^T, \\
\bm e_1^l := \frac{\sqrt{2}}{2} \left( \bm e_\alpha - \bm e_\beta \right), \quad
\bm e_2^l := \frac{\sqrt{2}}{2} \left( \bm e_\alpha + \bm e_\beta \right), \quad
\bm e_3^l := \bm n^f, \\
\bm e_\alpha := \frac12 \left( \bm e_\xi + \bm e_\eta \right) / \left \| \frac12 \left( \bm e_\xi + \bm e_\eta \right) \right \| , \quad
\bm e_\beta := \frac{\bm e_3^l \times \bm e_\alpha}{ \| \bm e_3^l \times \bm e_\alpha \| }.
\end{gather*}
The parameterization yields the following transformation of the volume element 
\begin{align*}
& d\Omega := h^s \bm n ^f \cdot \left(\frac{\partial \bm \chi}{\partial \xi} \times \frac{\partial \bm \chi}{\partial \eta} \right) d\xi d\eta d\zeta = h^s d\Gamma d\zeta,
\end{align*}
and consequently the following simplification of the volume integral over $\Omega^s$,
\begin{align}
\label{eq:shell-volume-integral}
\int_{\Omega^s} \left( \cdot \right) d\Omega 
= \int_{\Gamma_I} h^s \int_0^1 \left( \cdot \right) d\zeta d\Gamma.
\end{align}

\begin{assumption}
\label{assumption:membrane}
The displacement $\bm u^s$ is a function of the in-plane parametric coordinates $(\xi, \eta)$ only, and the transverse normal stress $\bm \sigma^s_{33}$ is zero in the $\bm e^l_3$ direction of the lamina system.
\end{assumption}
Cardiac pulse wavelengths are at least three orders of magnitude larger than arterial diameters, causing vessels to respond to transverse loading primarily with in-plane stresses. Out-of-plane rotations and their corresponding bending deformations are thus neglected under this membrane assumption, thereby minimizing the number of degrees of freedom and further facilitating convenient fluid-solid coupling. In addition, the transverse normal stress is assumed to vanish in order to avoid thickness locking \cite{Bischoff2004}. Assumption \ref{assumption:membrane} enables evaluation of $\left( \cdot \right)$ in \eqref{eq:shell-volume-integral} at $\zeta = 0$, thereby reducing the volume integral over $\Omega^s$ to a surface integral over $\Gamma_I$,
\begin{align*}
\int_{\Omega^s} \left( \cdot \right) d\Omega \approx \int_{\Gamma_I} h^s \left( \cdot \right)|_{\zeta = 0} d\Gamma.
\end{align*}

Considering an isotropic linear elastic solid material, the solid constitutive relation, expressed in the lamina coordinate system to enforce the zero transverse normal stress condition, is as follows,
\begin{align*}
& \bm \sigma^{s, l} = \bm \sigma^{s, l}_{\mathrm{dev}} - p^s \bm I = \mathbb C^{s, l}  \bm \epsilon^{l}(\bm u^{s,l}), \\ 
& \mathbb C^{s, l} := 2 \mu^s(\bm x^l) \mathbb I + \lambda^s(\bm x^l) \bm I \otimes \bm I,
\end{align*}
where $\bm \sigma^{s, l}$ and $\bm \epsilon^{l}(\bm u^{s,l})$ are respectively the Cauchy stress and infinitesimal strain in the lamina coordinate system, $\bm I$ is the second-order identity tensor, $\mathbb I$ is the fourth-order symmetric identity tensor, and $\mu^s$ and $\lambda^s$ are the Lam\'e parameters related to the bulk modulus in the form $\kappa^s := 2\mu^s/3 + \lambda^s$. In Voigt notation,
\begin{align*}
&\bm \sigma^{s, l} = \left[ \sigma_{11}^{s,l}, \sigma_{22}^{s,l}, \sigma_{12}^{s,l}, \sigma_{23}^{s,l} , \sigma_{31}^{s,l} \right]^T, \\
&\bm \epsilon^{l}(\bm u^{s, l}) = \left[ \epsilon_{11}^{l}, 
\epsilon_{22}^{l}, 2 \epsilon_{12}^{l}, 2 \epsilon_{23}^{l}, 2 \varepsilon_{31}^{l} \right]^T\nonumber \\ 
& \hspace{10mm} = \left[ u_{1,1}^{s,l} , u_{2,2}^{s,l}, u_{1,2}^{s,l} + u_{2,1}^{s,l}, u_{3,2}^{s,l}, u_{3,1}^{s,l} \right]^T, \\
& \mathbb C^{s, l} = \frac{E}{(1 - \nu^2)}
\begin{bmatrix}
1 & \nu &  & &  \\[1mm]
\nu & 1 & &  &  \\[1mm]
 &  & \displaystyle \frac{1 - \nu}{2} & &  \\[1mm]
 &  &  & \kappa \displaystyle \frac{(1 - \nu)}{2} &  \\[1mm]
 &  &  &  & \kappa \displaystyle \frac{ (1 - \nu)}{2} \\[1mm]
\end{bmatrix},
\end{align*}
where $E$ is the Young's modulus, $\nu$ is the Poisson's ratio, and $\kappa = 5/6$ is the shear correction factor \cite{Hughes1987}. We note the addition of transverse shear modes to enhance the linear membrane under transverse loads in three-dimensional structures. The Cauchy stress in the lamina coordinate system is then rotated to the global coordinate system by
\begin{align*}
\bm \sigma^s = \bm Q^T \bm \sigma^{s,l} \bm Q.
\end{align*}

The semi-discrete fluid formulation is constructed with the residual-based VMS formulation. Let $\mathcal S_{\bm v}^f$ and $\mathcal S_{p}^f$ denote the trial solution spaces for the fluid velocity and pressure; let $\mathcal V_{\bm v}^f$ and $\mathcal V_{p}^f$ be their corresponding test function spaces; and let $\Gamma_h^f$ denote the Neumann part of the fluid boundary with traction $\bm h^f$ prescribed. We can then state the weak form problem as follows. Find 
$\bm y_h^f(t):= \left\lbrace \bm v_h^f(t), p_h^f(t) \right\rbrace \in \mathcal S_{\bm v}^f \times \mathcal S_{p}^f$ such that $\forall \left\lbrace \bm w_h^f, q_h^f\right\rbrace \in \mathcal V_{\bm v}^f \times \mathcal V_{p}^f$,
\begin{align*}
& \mathbf B^{f}_{\mathrm{m}} \left( \bm w_h^f ;  \dot{\bm y}_h^f, \bm y_h^f \right) := \int_{\Omega^f} \bm w_h^f \cdot \rho^f \left( \frac{\partial \bm v_h^f}{\partial t} + \bm v_h^f \cdot \nabla \bm v_h^f - \bm b^f \right) d\Omega \\ 
& - \int_{\Omega^f} \nabla \cdot \bm w_h^f p_h^f d\Omega + \int_{\Omega^f} 2\mu^f  \bm \varepsilon(\bm w_h^f) : \bm \varepsilon(\bm v_h^f)  d\Omega \\
& - \int_{\Omega^{f \prime}} \nabla \bm w_h^f : \left( \rho^f \bm v^{\prime} \otimes \bm v_h^f \right)  d\Omega + \int_{\Omega^{f \prime}} \nabla \bm v_h^f : \left( \rho^f \bm w_h^f \otimes \bm v^{\prime} \right) d\Omega \\ 
& - \int_{\Omega^{f \prime}} \nabla \bm w_h^f : \left( \rho^f \bm v^{\prime} \otimes \bm v^{\prime} \right) d\Omega - \int_{\Omega^{f \prime}} \nabla \cdot \bm w_h^f p^{\prime} d\Omega \\
& - \int_{\Gamma_h^f} \bm w_h^f \cdot \bm h^f d\Gamma - \int_{\Gamma_h^f}  \rho^f \beta \left(\bm v_h^f \cdot \bm n^f \right)_{-} \bm w_h^f \cdot \bm v_h^f d\Gamma = 0, \\
& \mathbf B^{f}_{\mathrm{c}}\left( q_h^f; \dot{\bm y}_h^f, \bm y_h^f \right) := \int_{\Omega^f} q_h^f \nabla \cdot \bm v_h^f d\Omega  - \int_{\Omega^{f \prime}} \nabla q_h^f \cdot  \bm v^{\prime} d\Omega = 0, \\
& \bm v^{\prime} := -\bm \tau_{M} \left( \rho^f \frac{\partial \bm v_h^f}{\partial t} + \rho^f \bm v_h^f \cdot \nabla \bm v_h^f  + \nabla p_h^f - \mu^f \Delta \bm v_h^f - \rho^f \bm b^f \right), \displaybreak[2] \\
& p^{\prime} := -\tau_C \nabla \cdot \bm v_h^f,
\end{align*}
wherein
\begin{align*}
& \tau_M := \frac{1}{\rho^f}\left( \frac{\mathrm C_{\mathrm T}}{\Delta t^2} + \bm v_h^f \cdot \bm G \bm v_h^f + \mathrm C_{\mathrm I} \left( \frac{\mu^f}{\rho^f} \right)^2 \bm G : \bm G \right)^{-\frac12}, \displaybreak[2] \\
& \tau_C := \frac{1}{\tau_M \textup{tr}\bm G}, \displaybreak[2] \\
& G_{ij} := \sum_{k=1}^{3} \frac{\partial y_k}{\partial x_i} M_{kl} \frac{\partial y_l}{\partial x_j}, \quad \bm M = [ M_{kl} ] = \frac{\sqrt[3]{2}}{2}\begin{bmatrix}
2 & 1 & 1 \\
1 & 2 & 1 \\
1 & 1 & 2
\end{bmatrix}, \displaybreak[2] \\
& \bm G : \bm G := \sum_{i,j=1}^{3} G_{ij} G_{ij}, \quad \textup{tr}\bm G := \sum_{i=1}^{3} G_{ii}, \displaybreak[2] \\
& \left( \bm v_h^f \cdot \bm n^f \right)_{-} := \frac{\bm v_h^f \cdot \bm n^f - |\bm v_h^f \cdot \bm n^f|}{2}.
\end{align*}
Here, $\bm y = \left\lbrace y_i \right\rbrace_{i=1}^{3}$ are natural coordinates in the parent domain, and $C_I$ and $C_T$ are taken to be $36$ and $4$ in this study. $\bm M$ is introduced to yield node-numbering-invariant definitions of $\tau_M$ and $\tau_C$ for simplex elements \cite{Danwitz2019}. The final term in $\mathbf B^{f}_{\mathrm{m}}$ is an additional convective traction shown to be robust in overcoming backflow divergence \cite{Bazilevs2009b, Moghadam2011}, a well-known issue in cardiovascular simulations. It can be shown that taking $\beta = 1.0$ guarantees energy stability for the numerical scheme adopted here. In this work, $\beta$ is fixed to be $0.2$ to minimize its impact on the flow field and to improve robustness at larger time steps. 

Discretization of the entire domain $\Omega$ by a single mesh with continuous basis functions across the fluid-solid interface $\Gamma_I$ immediately guarantees satisfaction of the kinematic coupling condition $\bm v^f = \bm v^s$ in the semi-discrete formulation. The implied relation $\bm w^f_h = \bm w^s_h$ on $\Gamma_I$ also yields weak satisfaction of 
the traction coupling condition, that is 
\begin{align*}
0 = \int_{\Gamma_I} \bm w^f_h \cdot \left( \bm \sigma^f \bm n^f + \bm \sigma^s \bm n^s \right) d\Gamma.
\end{align*}
With this mesh choice, the momentum balances over $\Omega^f$ and $\Omega^s$ can then be combined into a single momentum balance over $\Omega$, 
\begin{align*}
\mathbf B^{s}_{\mathrm{m}}\Big( \bm w^s_h ;  \dot{\bm y}^s_h, \bm y^s_h \Big) + \mathbf B^{f}_{\mathrm{m}} \left( \bm w^f_h ;  \dot{\bm y}^f_h, \bm y^f_h \right) = 0.
\end{align*}
Having applied the outlined assumptions to collapse the three-dimensional elastodynamic problem in $\Omega^s$ to a two-dimensional problem posed on $\Gamma_I$, we can now present the reduced semi-discrete FSI formulation. Let $\bm u^{w}_h$ be the membrane displacement on $\Gamma_I$. Using the kinematic coupling condition, continuity of test functions on $\Gamma_I$, and the transformation of volume integrals over $\Omega^s$, we can rewrite the kinematic equation in $\Omega^s$ as
\begin{align*}
& \mathbf B_{\mathrm{k}} \left( \dot{\bm y}_h, \bm y_h \right) := \frac{d\bm u^w_h}{dt} - \bm v^f_h = \bm 0, \qquad \mbox{ on } \Gamma_I.
\end{align*}
and the momentum balance over $\Omega^s$ as 
\begin{align*}
& \mathbf B^w_{\mathrm{m}} \left( \bm w_h^f ;  \dot{\bm y}_h, \bm y_h \right) := \int_{\Gamma_I} \bm w_h^f \cdot \rho^s h^s \left( \frac{d \bm v_h^f}{d t} - \bm b^s \right) d\Gamma \nonumber \\
& + \int_{\Gamma_I} h^s \bm \epsilon(\bm w_h^f) : \bm \sigma^s(\bm u_h^w)  d\Gamma - \int_{\partial \Gamma_I \cap \Gamma^h_s } h^s \bm w_h^f \cdot \bm h^s d\Gamma,
\end{align*}
where $\partial \Gamma_I \cap \Gamma^h_s$ constitutes the Neumann partition of the boundary of $\Gamma_I$. Finally, let $\mathcal S^{w}_{\bm u}$ be the trial solution space for the membrane displacement on $\Gamma_I$. Our RUC formulation posed only in $\Omega^f$ is then stated as follows. Find $\bm y_h(t) := \left\lbrace \bm u^{w}_h(t), \bm v_h^f(t), p_h^f(t) \right\rbrace \in \mathcal S^w_{\bm u} \times \mathcal S_{\bm v}^f \times \mathcal S_{p}^f$ such that
\begin{align*}
& \mathbf B_{\mathrm{k}} \left( \dot{\bm y}_h, \bm y_h \right) = \bm 0, &&  \\
& \mathbf B_{\mathrm{m}} \left( \bm w_h^f ;  \dot{\bm y}_h, \bm y_h \right) := \mathbf B^w_{\mathrm{m}} \left( \bm w_h^f ;  \dot{\bm y}_h, \bm y_h \right) + \mathbf B^f_{\mathrm{m}} \left( \bm w^f_h ;  \dot{\bm y}^f_h, \bm y^f_h \right)  = 0, \\
& \mathbf B_{\mathrm{c}}\left( q^f_h; \dot{\bm y}_h, \bm y_h \right) := \mathbf B^f_{\mathrm{c}}\left( q^f_h; \dot{\bm y}^f_h, \bm y^f_h \right) = 0,
\end{align*}
$\forall \left\lbrace \bm w_h^f, q_h^f\right\rbrace \in \mathcal V_{\bm v}^f \times \mathcal V_{p}^f$.

\subsection{Solution strategy}
The generalized-$\alpha$ method is applied for temporal discretization of the semi-discrete FSI formulation derived in Section \ref{subsec:semi-discrete-fsi}, in which both velocity and pressure are collocated at the intermediate time step to achieve uniform second-order temporal accuracy \cite{Liu2020a}. Without loss of consistency, the fully discrete scheme is solved with a segregated predictor multi-corrector algorithm preserving the two-by-two block structure of the incompressible Navier-Stokes equations in the implicit solver's associated linear system. In particular, only the upper left block matrix associated with the momentum equations and velocity degrees of freedom is modified to include a wall contribution. Block preconditioning of a monolithically coupled FSI system is therefore made possible, and the membrane displacement is simply updated algebraically in each nonlinear iteration. For improved representation of the Schur complement, we apply the nested block preconditioner \cite{Liu2020} that algorithmically defines the action of the Schur complement on a vector in a matrix-free fashion. We have demonstrated enhanced robustness and scalability of our block preconditioner as compared to alternative preconditioners in applications spanning hyperelasticity, viscous fluids, and FSI \cite{Liu2019}.

\subsection{Tissue prestressing}
The semi-discrete FSI formulation above assumes the in vivo vascular wall configuration at imaging to be stress-free, yet an internal stress state, termed the prestress, must exist to balance the in vivo blood pressure and viscous traction. In contrast to approaches that seek to determine a stress-free configuration \cite{Tezduyar2008, Nama2020}, we instead generate the prestress $\bm \sigma_0$ via a fixed-point algorithm similar to the one proposed for an ALE formulation \cite{Hsu2011}, in which we consider the following variational problem for the vascular wall. Given the body force per unit mass $\bm b^s$, boundary traction $\bm h^s$, and fluid boundary traction $\bm h^f$, find $\bm u_h^w \in \mathcal S^w_{\bm u}$ and $\bm v^w_h \in \mathcal S^w_{\bm v}$, such that $\forall \bm w^f_h \in \mathcal{V}^f_{\bm v}$,
\begin{align}
\label{eq:semi-discrete-prestress-kinematics}
& \bm 0 = \frac{d \bm u^w_h}{dt} - \bm v^w_h, \\
\label{eq:semi-discrete-prestress-u}
& 0 = \mathbf B^w_{\mathrm{m}} \left( \bm w_h^f ;  \dot{\bm y}_h, \bm y_h\right) + \int_{\Gamma_I} \bm w_h^f \cdot \bm h^f d\Gamma,
\end{align}
where
\begin{align*}
& \mathbf B^w_{\mathrm{m}} \left( \bm w_h^f ;  \dot{\bm y}_h, \bm y_h\right) := \int_{\Gamma_I} \bm w_h^f \cdot \rho^s h^s \left( \frac{d \bm v_h^f}{d t} - \bm b^s \right) d\Gamma  \nonumber \\
& + \int_{\Gamma_I} h^s \bm \epsilon(\bm w_h^f) : \Big( \bm \sigma^s(\bm u_h^w) + \bm \sigma_0 \Big) d\Gamma - \int_{\partial \Gamma_I \cap \Gamma^h_s } h^s \bm w_h^f \cdot \bm h^s d\Gamma,
\end{align*}
and $\mathcal S^w_{\bm v}$ is a suitable trial solution space for the wall velocity. Using the prestress generation algorithm summarized below, $\bm \sigma_0$ is then determined such that equations \eqref{eq:semi-discrete-prestress-kinematics}-\eqref{eq:semi-discrete-prestress-u} are satisfied under the imaged wall configuration. We denote the prestress at the $m$-th fixed-point iteration as $\bm \sigma_{0,(m)}$ and the maximum number of iterations as $m_{\mathrm{max}}$.

\begin{algorithm}[H]
\caption{Prestress generation algorithm}
\begin{algorithmic}[1]
\Statex \hskip-1.5em \textbf{Initialization:} Set $\bm \sigma_{0, (0)} = \bm 0$, $\bm v^w_{0}=\bm 0$, and $\bm u^w_{0} = \bm 0$.

\Statex \hskip-1.5em \textbf{Fixed-point iteration:} Repeat for $m=0, 1, ..., m_{\mathrm{max}}$.

\State Set $\bm \sigma_0 = \bm \sigma_{0, (m)}$, $\bm v^w_m = \bm 0$, and $\bm u^w_m = \bm 0$.

\State From $t_m$ to $t_{m+1}$, solve the variational problem \eqref{eq:semi-discrete-prestress-kinematics}-\eqref{eq:semi-discrete-prestress-u} for $\bm u^w_{m+1}$ and $\bm v^w_{m+1}$ using the backward Euler method for temporal discretization.

\State Update $\bm \sigma_{0, (m+1)} = \bm \sigma^s(\bm u^w_{m+1}) + \bm \sigma_{0, (m)}$.

\State Let $\mathrm{tol}_{\mathrm{P}}$ be a prescribed tolerance. If the stopping criterion $\| \bm u^w_{m+1} \|_{\mathfrak l_2} \leq \mathrm{tol}_{\mathrm{P}}$ is satisfied, then set $\bm \sigma_0 = \bm \sigma_{0,(m+1)}$ and exit the fixed-point iteration.

\end{algorithmic}
\end{algorithm}

\section{Patient-specific abdominal aortic aneurysm}
\label{sec:patient-specific-AAA}
Of the recent FSI studies of patient-specific AAAs \cite{DiMartino2001, Wolters2005, Rissland2009, Fonken2021}, none modeled the supraceliac aorta, as would be crucial when assuming an axisymmetric inlet velocity profile \cite{Les2010}, such as the parabolic and plug profiles assumed in these studies. Using the open-source software package SimVascular \cite{Lan2018,Updegrove2017}, we modeled a patient-specific AAA from the computed tomography angiogram of a $75$-year-old male, including a total of $11$ outlets between the supraceliac aorta and common iliac arteries. A linear tetrahedral mesh of $2.6 \times 10^6$ elements was generated with MeshSim (Simmetrix, Inc., Clifton Park, NY, USA) with three boundary layers at a thickness gradation factor of $0.5$.

Adopting centimeter-gram-second units, we set the fluid density $\rho^f$ to $1.06$, fluid viscosity $\mu^f$ to $0.04$, wall density $\rho^s$ to $1.0$, and wall Poisson's ratio $\nu$ to $0.5$. The wall thickness $h^s$ over the non-aneurysmal and aneurysmal regions was prescribed to be $11\%$ and $6\%$ of the local centerline-based radius, respectively, to achieve experimentally measured values \cite{VandeGeest2006}. Two different distributions for the wall Young's modulus $E$ were investigated based on the maximum tangential moduli derived from uniaxial \cite{DiMartino2006} and biaxial \cite{VandeGeest2006} tensile testing: a) aneurysmal $2.02 \times 10^7$, non-aneurysmal $6.73 \times 10^6$, b) aneurysmal $1.17 \times 10^8$, non-aneurysmal $3.90 \times 10^7$. For the remainder of our work, we refer to these Young's modulus distributions as \textit{E-uniaxial} and \textit{E-biaxial}. Given the unavailability of uniaxial testing-derived material properties for non-aneurysmal tissue, we selected the non-aneurysmal modulus in \textit{E-uniaxial} to achieve the same moduli ratio between the two types of tissue as in \textit{E-biaxial}. Of note, the \textit{E-biaxial} maximum tangential moduli correspond to the reported wall behavior in the circumferential direction. In addition to \textit{E-uniaxial} and \textit{E-biaxial}, we further investigated the commonly employed rigid wall assumption \cite{DiAchille2014, Soudah2013, Suh2011, Les2010a}. 

Identical boundary conditions were prescribed across the three sets of wall material properties. At the inlet, we prescribed a representative supraceliac aortic flow waveform with a parabolic velocity profile. Using a modular implicit method for 3D-0D coupling \cite{Moghadam2013}, three-element Windkessel models were coupled at the outlets and tuned to achieve patient-specific inlet systolic and diastolic pressures ($118$ / $78$ mm Hg) for \textit{E-uniaxial} as well as flow splits from the literature \cite{MooreJr.1994, Les2010}. Specifically, $66\%$ of the supraceliac aortic inflow was distributed to the upper branches, with the remaining $34\%$ continuing to the infrarenal aorta. Among the upper branch flows, $33\%$ was distributed to the celiac trunk with an even flow split to the hepatic and splenic arteries, and the remainder was evenly distributed to the superior mesenteric artery, left renal artery, and right renal arteries. Fourteen percent of the infrarenal flow was distributed to the inferior mesenteric artery, and the remainder was evenly split among the left and right common iliac arteries, with a $70\%$-$30\%$ external-internal iliac artery flow split on each side.

\begin{figure}
\begin{center}
\includegraphics[trim=90 115 110 94, clip=true, width=0.98\textwidth]{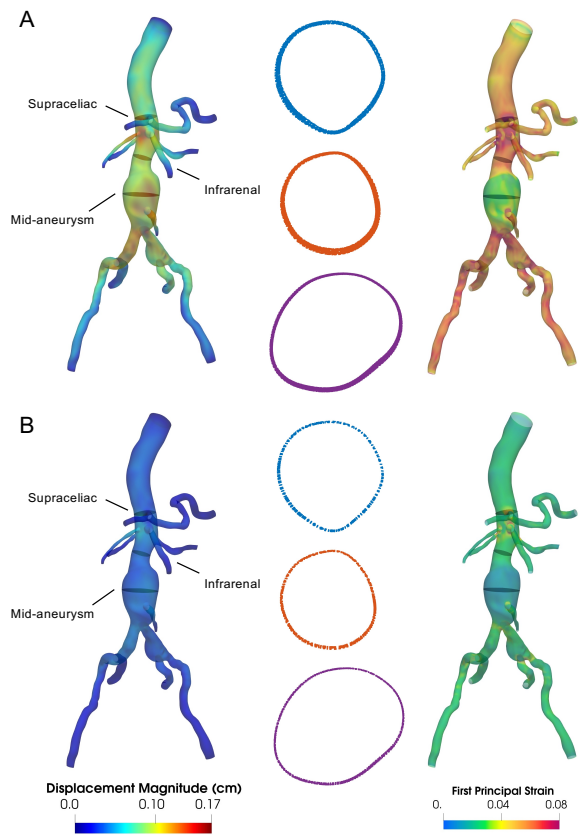}
\caption{Peak systolic wall displacement magnitudes, first principal Green-Lagrange strain, and overlaid wall node positions over time at supraceliac (blue), infrarenal (orange), and mid-aneurysm (purple) cross-sectional slices for the (A) \textit{E-uniaxial} and (B) \textit{E-biaxial} Young's modulus distributions.}
\label{fig:disp-strain-contours}
\end{center}
\end{figure}

Initial conditions were generated by first running a rigid-walled CFD simulation to generate solution fields at the diastolic pressure. For the two FSI simulations, we subsequently used the prestress generation algorithm to obtain the prestress $\bm \sigma_0$ balancing the diastolic fluid boundary traction under zero wall displacement relative to the imaged configuration. Simulations were performed over three cardiac cycles with uniform time steps and verified for convergence to a limit cycle. Only the final cardiac cycle was analyzed. Using three Intel Xeon Gold 5118 processors interconnected by a 100 GB/s EDR InfiniBand for a total of 72 threads operating at 191 GB RAM and a clock rate of 2.3 GHz, the \textit{E-uniaxial} and \textit{E-biaxial} simulations respectively required $1.3$X and $1.8$X the CPU time of the rigid wall simulation (7 hr 56 min). This difference in CPU time can be understood from the increasing heterogeneity in the linear system's upper left block matrix when moving from the rigid wall case to \textit{E-uniaxial} and finally to \textit{biaxial}, necessitating an increasing number of nonlinear iterations to converge.

\begin{figure}
\begin{center}
\includegraphics[trim=32 90 140 95, clip=true, width=1.4\textwidth]{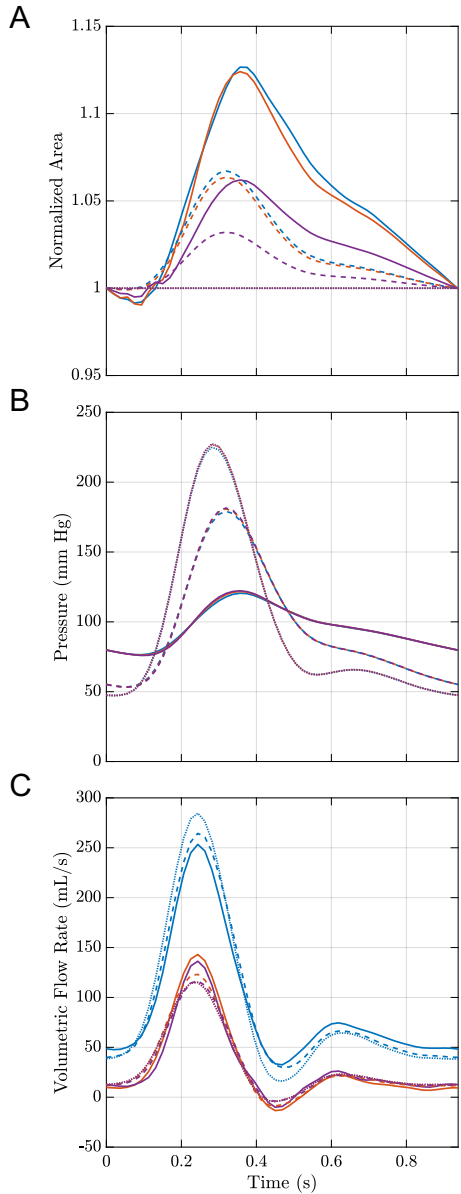}
\caption{Normalized cross-sectional areas, spatially averaged pressures, and volumetric flow rates over time at the supraceliac (blue), infrarenal (orange), and mid-aneurysm (purple) slices for \textit{E-uniaxial} (solid), \textit{E-biaxial} (dashed), and rigid wall (dotted). Normalization was performed with respect to the corresponding cross-sectional areas at time $t=0$.}
\label{fig:slice-area-pres-flow}
\end{center}
\end{figure}

\begin{figure}
\begin{center}
\includegraphics[trim=70 270 110 95, clip=true, width=1.0\textwidth]{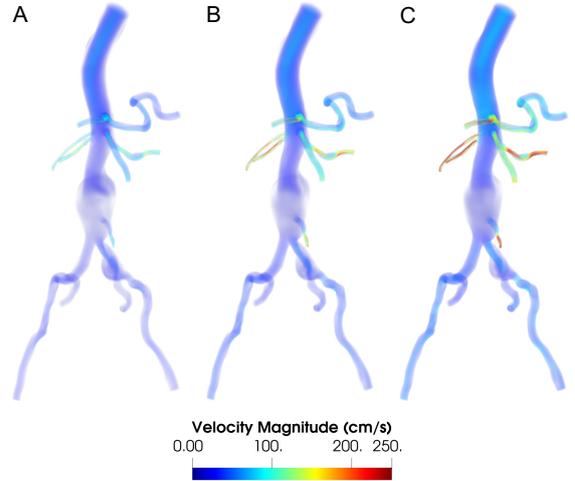}
\caption{Volume rendering of peak systolic velocity magnitude for (A) \textit{E-uniaxial}, (B) \textit{E-biaxial}, and (C) rigid wall.}
\label{fig:velocity-rendering}
\end{center}
\end{figure}

The significantly stiffer maximum tangential moduli of \textit{E-biaxial} yields a maximum peak systolic wall displacement magnitude over $2.4$X smaller than that of \textit{E-uniaxial} (Figure \ref{fig:disp-strain-contours}). The first principal Green-Lagrange strain averaged over the aneurysm sac at peak systole is also approximately $2.4$X smaller. We note that whereas the \textit{E-uniaxial} strain of 0.0344 falls within a previously reported range for human AAAs ($0.032$ -- $0.091$) based on in vivo cine magnetic resonance imaging \cite{Satriano2015}, the \textit{E-biaxial} strain of 0.0146 is much too small. To facilitate comparisons with existing luminal wall motion data from cardiac-gated 2D cine gradient-echo magnetic resonance imaging \cite{Les2010a}, we overlaid all wall nodes over time at three cross-sectional slices along the abdominal aorta: supraceliac (SC), infrarenal (IR), and mid-aneurysm (mid-AAA). Asymmetry can be observed in these temporal overlays, with the aortic wall expanding more on the anterior side. Cross-sectional areas at these same slices, normalized by their respective values at time $t=0$, are plotted over time in Figure \ref{fig:slice-area-pres-flow}A. Assuming circular profiles, maximum effective diameter changes for \textit{E-uniaxial} ($6.14\%$ SC, $6.02\%$ IR, $3.05\%$ mid-AAA) are in excellent agreement with those reported as cohort averages in \cite{Les2010a} ($6.69\%$ SC, $5.34\%$ IR, $2.60\%$ mid-AAA) as well as maximum aneurysmal displacements measured from dynamic computed tomography \cite{Piccinelli2013}. In contrast, the corresponding maximum effective diameter changes for \textit{E-biaxial} ($3.30\%$ SC, $3.13\%$ IR, and $1.59\%$ mid-AAA) are 49\% to 61\% of the cohort averages. As is evident from the spatially averaged pressure profiles over time (Figure \ref{fig:slice-area-pres-flow}B), decreases in 3D capacitance and thus decreases in total capacitance yield increased pressure pulses (\textit{E-uniaxial} $40.3$ mm Hg, \textit{E-biaxial} $122$ mm Hg, rigid wall $176$ mm Hg). Furthermore, the phase delay of $0.0375$ s and $0.0750$ s from peak systole in the rigid wall simulation to that in \textit{E-uniaxial} and \textit{E-biaxial}, respectively, is reflective of the increased wave speed in stiffer vessels. Despite consistent mean flow distributions across the three simulations (Figure \ref{fig:slice-area-pres-flow}C), volume-rendered velocity magnitudes depict the increased velocities at peak systole for \textit{E-biaxial} and rigid wall relative to \textit{E-uniaxial} (Figure \ref{fig:velocity-rendering}), a combined effect of the increased flow amplitudes and reduced area changes. While Lin et al. \cite{Lin2017} noted that the rigid wall assumption overestimates the wall shear stress (WSS) at the aneurysm neck and underestimates WSS in the aneurysm sac, we instead observed similar OSI ($0.163$ -- $0.180$) and TAWSS ($3.15$ -- $3.39$ dyn/cm$^2$) values averaged over the aneurysm sac across all three cases. These results are in agreement with the similar velocity profiles and WSS gradients observed across rigid wall and FSI models in \cite{Wolters2005}.

\begin{figure}
\begin{center}
\includegraphics[trim=60 95 150 95, clip=true, width=1.0\textwidth]{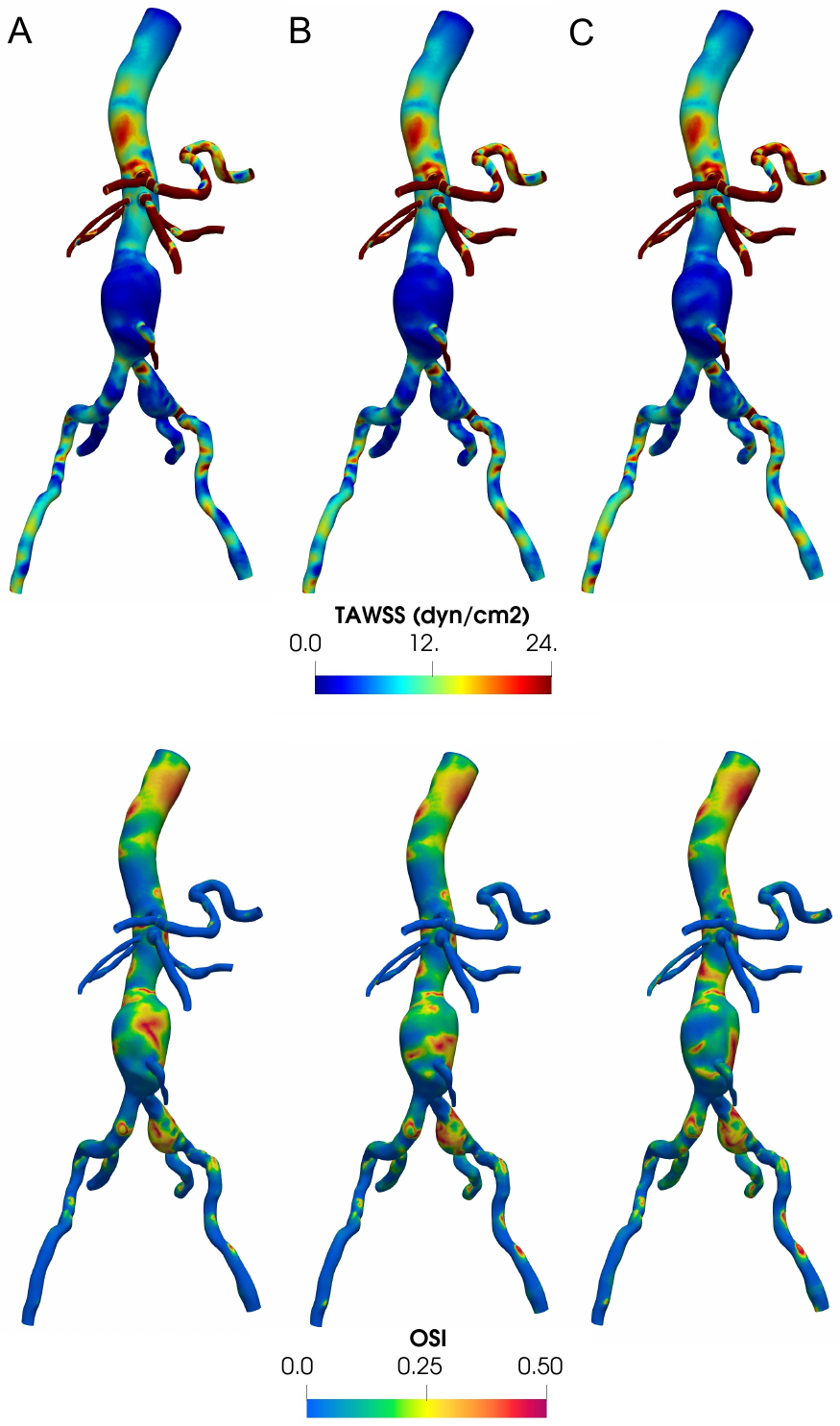}
\caption{Time-averaged wall shear stress (TAWSS, top) and oscillatory shear index (OSI, bottom) for (A) \textit{E-uniaxial}, (B) \textit{E-biaxial}, and (C) rigid wall.}
\label{fig:tawss-osi}
\end{center}
\end{figure}

\section{Conclusions}
We have presented a computationally efficient framework for patient-specific vascular FSI modeling, involving our RUC formulation and in vivo tissue prestressing. In reducing the unified continuum formulation in the ALE description, the small-strain assumption guarantees geometry adherence and eliminates the need for mesh motion; the thin-walled assumption collapses the 3D structural problem to a 2D problem posed on the fluid-solid interface; and the membrane assumption facilitates convenient fluid-solid coupling via the same velocity degrees of freedom on the fluid-solid interface. The resulting monolithically coupled FSI formulation in the Eulerian frame differs from the semi-discrete fluid formulation only in the fluid boundary integral, thus requiring CPU times only minimally more expensive than rigid wall formulations. This computational efficiency is particularly significant in the context of hemodynamically-driven growth and remodeling simulations requiring frequent updates in both WSS and wall strain.

We additionally emphasize novel aspects of our numerical strategy pertaining to the temporal discretization and linear solver. Temporal discretization of the entire FSI system is performed with the generalized-$\alpha$ scheme, in which velocity and pressure are uniformly evaluated at the intermediate time step to achieve uniform second-order temporal accuracy, in direct contrast to the predominant dichotomous approach offering only first-order accuracy of pressure \cite{Liu2020a}. Furthermore, block preconditioning of a monolithically coupled FSI system is made possible for the first time through a segregated predictor multi-corrector algorithm preserving the block structure of the incompressible Navier-Stokes equations in the implicit solver's fully consistent linear system. 

Despite unavailability of time-resolved angiography for validation of our patient-specific AAA investigation, we achieved excellent agreement of our predicted wall motion at multiple locations along the abdominal aorta with existing imaging studies when prescribing wall properties derived from uniaxial tensile testing \cite{Les2010a, Piccinelli2013, Satriano2015}. The large discrepancy in maximum tangential moduli between \textit{E-uniaxial} and \textit{E-biaxial} can be understood from the predicted mechanical responses from the two testing protocols \cite{DiMartino2006, VandeGeest2006}. While both nonlinear, the uniaxial response of AAA tissue is much stiffer than the biaxial response in the low strain (under 12\%) regions but more compliant in the high-strain regions. Whereas uniaxial loading allows for earlier recruitment and alignment of collagen fibers along a single loading axis and thus exhibits less nonlinearity, the orthogonally applied load in biaxial tension yields a much longer toe region in the biaxial response, after which the tissue becomes markedly stiffer \cite{VandeGeest2006}. Given the maximum tangential moduli-based linear constitutive model adopted in our study and the low strains characteristic of AAA, the superior performance of the \textit{E-uniaxial} Young's modulus distribution is unsurprising. Looking ahead, our RUC formulation could readily be extended to anisotropic nonlinear structural models. Future investigations involving larger patient cohorts could be validated against blood velocity and wall motion data from time-resolved phase contrast magnetic resonance imaging and/or computed tomography.

\section*{Acknowledgements}
This work was supported by the National Institutes of Health [grant numbers 1R01HL121754, 1R01HL123689, R01EB01830204], Southern University of Science and Technology [startup grant number Y01326127], the National Natural Science Foundation of China [grant number 12172160], and the Guangdong-Hong Kong-Macao Joint Laboratory for Data-Driven Fluid Mechanics and Engineering Applications [grant number 2020B1212030001]. Ingrid S. Lan was supported by the National Science Foundation (NSF) Graduate Research Fellowship and the Stanford Graduate Fellowship in Science and Engineering. Computational resources were provided by the Stanford Research Computing Center, the Extreme Science and Engineering Discovery Environment \cite{Towns2014} supported by NSF [grant number ACI-1053575], and the Center for Computational Science and Engineering at Southern University of Science and Technology.

\bibliographystyle{elsarticle-num}
\bibliography{forces-in-mechanics}

\end{document}